\documentclass[aps,prl,twocolumn]{revtex4}
\usepackage{graphicx}
\usepackage{bm}
\usepackage{amsmath}
\usepackage[colorlinks=true,allcolors=blue]{hyperref}

\providecommand{\bra}[1]{\langle #1 \rvert}
\providecommand{\ket}[1]{\lvert #1 \rangle}

\providecommand{\be}{\begin{equation}}
\providecommand{\ee}{\end{equation}}
\providecommand{\ba}{\begin{eqnarray}}
\providecommand{\ea}{\end{eqnarray}}

\usepackage{mathbbol}
\usepackage[makeroom]{cancel}

\begin{document}

\title{Investigating nonclassicality in nonlinear electronic spectroscopy}

\author{Saulo V. Moreira and Fernando L. Semi\~ao}

\affiliation{Centro de Ci\^encias Naturais e Humanas, Universidade Federal do ABC - UFABC, Santo Andr\'e, Brazil}

\begin{abstract}
In this work, we establish a connection between nonlinear electronic spectroscopy and the protocol for the non-disturbance condition, the non-fullfilment of which is a witness of nonclassicality  and can be related to the presence of coherence.
 Our approach permits us to express the nonclassicality witness condition in terms of common observables in the context of electronic spectroscopy experiments, such as the induced polarization. 
In this way, we provide the theoretical framework allowing one to infer nonclassicality from the detected signals in these experiments.
 

\end{abstract}
\vskip2pc

\maketitle

\section{I. Introduction}

Nonlinear spectroscopy techniques such as two-dimensional electronic spectroscopy are well-established both from a theoretical and experimental viewpoint \citep{Mukamel}. 
Among its most noteworthy applications, one could mention its usefulness to reveal electronic couplings and energy transfer pathways in complex molecular aggregates \citep{pathways}. Consequently, these experimental techniques assume a prominent role in the development of the field of \textit{quantum biology}  \citep{Lambert}. 
Within this field, a considerable amount of investigation has been undertaken in view of determining whether nonclassical features such as quantum coherence may contribute to optimizing the energy transfer within multichromophoric systems. A paradigmatic example is the so-called Fenna-Matthew-Olson (FMO) complex present in green sulfur bacteria \citep{Engel, Ishizaki, Panitchayangkoon, Collini}. 
The approaches usually  rely on particular theoretical models or are focused on the oscillations of the cross peaks in the 2D electronic spectroscopy signals as a signature of nonclassicality. 
Nonetheless, alternative ways of witnessing nonclassical effects within these systems, in particular the ones which do not specify a particular model, may permit us to have a clearer and more comprehensive picture of the different aspects of nonclassicality in molecular systems. 
As a result, we could push the boundaries of the field of quantum biology a little step further.


Classical physics assumes {\it measurement noninvasiveness}, which is to say that, in principle, one can perform measurements on classical systems with arbitrarily small disturbance on their dynamics. Conversely, one aspect of the nonclassicality of physical systems is \textit{invasiveness}, meaning that measurements on these systems may affect their subsequent evolution in such a way that the statistics of later measurements performed on the same system cannot be explained by classical theory.
Measurement noninvasiveness can be ruled out through the violation of an inequality proposed by Leggett and Garg in 1985 \cite{LG, Emary}. 
Specifically, the Leggett-Garg inequality (LGI) consists of a sum of two-time temporal quantum correlations between the results of dichotomic measurements performed on a single physical system as it evolves in time.
Originally, the LGI derivation was based on an assumption called ``macrorealism {\it per se}'' as well, in addition to non-invasive measurability. By macrorealism {\it per se} one means  the intuition that macroscopic objects -- here taken as objects obeying classical laws -- must be in a definite state at all instants of time, i.e. superpositions are not allowed.
 The two aforementioned assumptions are the content of the original definition of macrorealism by Leggett and Garg.

Many recent works have further developed and discussed Leggett and Garg original proposition, as well as the meaning of the violation of the inequality \cite{Maroney, Moreira, Clemente, Clemente2, Kumari, Knee2, Kofler2,Martin,Animesh}. Alternative definitions and witnesses to macrorealism have also been proposed in relation with the Leggett-Garg scenario.
For instance, the definition of the \textit{no-signaling in time}  (NSIT) condition \citep{Kofler2} puts forward the macrorealist assumption related to the null effect that the measurement process is expected to have on a physical system, i.e. measurement noninvasiveness. 
In this context, measurement invasiveness can be defined as the nonclassical effect of measurements on the statistics of another observable measured later.
Similar conditions to the NSIT have been used in Ref. \citep{Maroney} to derive the LGI, allowing one to conclude that measurement noninvasiveness can always be dismissed by the LGI violation.

By considering the definition of the non-disturbance condition \cite{Knee2,Schild,Lambert2}, we explore here the extended concept of \textit{noninvasiveness of an operation}, where the concept of measurement noninvasiveness is generalized to the notion of noninvasiveness of a general quantum channel  \cite{NewKnee, Moreira4}.
Invasiveness of an operation is aimed to detect nonclassical effects on the statistics of an observable measured at a later time, in the same spirit as the NSIT and Leggett-Garg scenarios  \cite{Knee2}. 
Within this framework, we propose a protocol for testing the non-disturbance condition in the context of electronic spectroscopy experiments, which could be employed with the purpose of witnessing nonclassicality in a vast range of systems wherein electronic spectroscopy is a tremendously useful tool.

\begin{figure}[h!]
\centering 
\includegraphics[width=8.6cm]{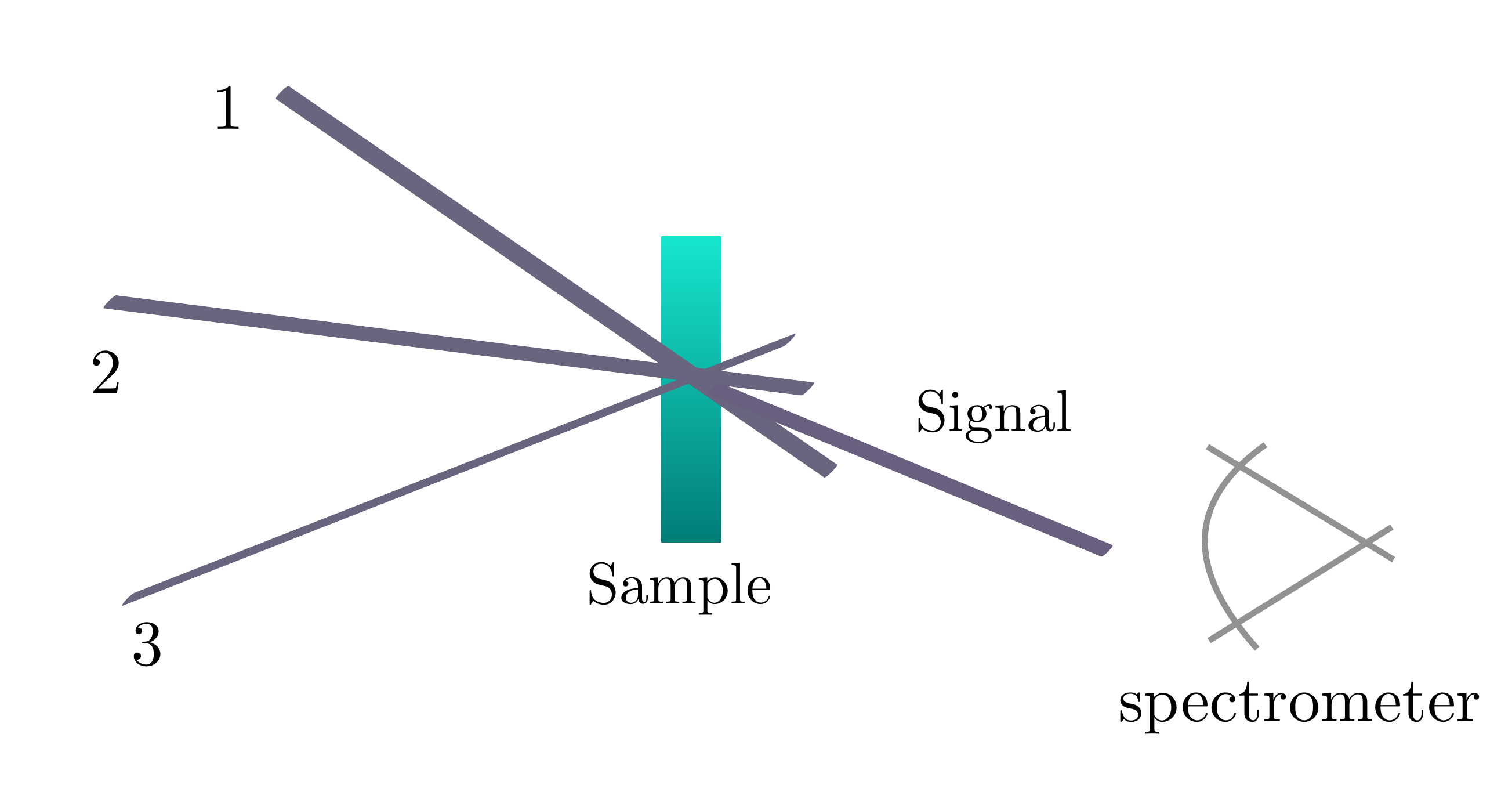}
\label{fig:fig14}
\caption{(Color online) Scheme of a 2D electronic spectroscopy experiment.  The laser pulses $1,\,2,$ and $3$ are labeled according to the order they are applied. The center of the pulses reach the sample at $t=0$, $t=t^\prime_1$, and  $t=t^\prime_2$, respectively. The time intervals between the interactions are $t^\prime_1-0=t_1$ and $t^\prime_2-t^\prime_1=t_2$, respectively. The time interval between the detection of the signal field by the spectrometer at time $t$ and pulse $3$ is $t_3$: $t-t^\prime_2=t_3$. \label{2DES}}
\end{figure}

\section{II. Nonlinear Response}

In typical two-dimensional electronic spectroscopy experiments, three laser pulses, labelled as pulse 1, 2 and  3, interact with the sample as sketched in Fig. \ref{2DES}. 
The quantum mechanical description of the experiment considers a time-dependent Hamiltonian, which can be expressed as
\begin{equation}\label{Ht}
H(t) = H_0 + H^\prime(t),
\end{equation}
where $H_0$ is time-independent and concerns the molecular system only, while $H^\prime(t)$ is  time-dependent and represents the action of the pulses. Before the application of the first pulse, the system is found to be in an equilibrium state $\rho_0$ i.e., a convex combination of the eigenstates of $H_0$. By denoting the electric field at time $t$ by $E(t)$, $H^\prime(t)$ is given by $H^\prime(t)=\mu\cdot E(t)$, where $\mu$ is the molecule dipole operator.  Quite important now is the fact that usually the sample-pulse interaction is weak enough to permit the safe use of perturbation theory \citep{Mukamel}. Within this approach, the following physical picture emerges.

When the first pulse reaches the sample, at time $t=0$, it potentially takes the system out of equilibrium, and coherence can be generated. 
Then, the system continues to undergo a free evolution governed by the Hamiltonian $H_0$ from $t=0$ to $t^\prime_1$, when the second pulse is applied.
Once again the molecular system is let to evolve freely, until the last interaction takes place at $t^\prime_2$. The sample-emitted electric field $E_S$ is detected at time $t$.  More precisely, what is detected is the intensity of the signal $I_S$ which is proportional to the electric field modulus squared  $I_S \propto |E_S|^2$. The source of such emitted field is the induced third-order nonlinear polarization $P^{(3)}(t)$ resulting from the application of the three pulses. A simple textbook calculation leads to \citep{Mukamel}
\begin{multline}\label{nonlinearP}
P^{(3)}(t) = \int_0^\infty dt_3 \int_0^\infty dt_2 \int_0^\infty dt_1 E(t-t_3) \\
\times E(t-t_3-t_2)E(t-t_3-t_2-t_1)S^{(3)}(t_3,t_2,t_1),
\end{multline}
where $S^{(3)}(t_1,t_2,t_3)$ is the third-order response function which reads
$
S^{(3)}(t_1,t_2,t_3) = \langle \mu(t_3+t_2+t_1)[\mu(t_2+t_1),[\mu(0),\rho(0)]] \rangle,
$
with $\mu(t)=U^\dagger_0(t)\mu U_0(t)$. Here,  $t^\prime_1-0=t_1$ is the time elapsed between the first and second pulses, $t^\prime_2-t^\prime_1=t_2$, the same for second and third pulses, and $t-t^\prime_2=t_3$, the time elapsed between the last pulse the signal detection. After the rotating wave approximation and Fourier decomposition of the third-order nonlinear polarization as 
\begin{equation}\label{FourierAmpl}
P^{(3)}(t) = \sum_s P_s^{(3)}(t) e^{i {\bf k_s}\cdot {\bf r} - i\omega_s t}, 
\end{equation}
where  ${\bf k_s=\pm k_1 \pm k_2 \pm k_3}$ and $\omega_s=\pm \omega_1 \pm \omega_2 \pm \omega_3$, the choice of one particular Fourier component in (\ref{FourierAmpl}) (phase matching) leads to 
$I_S(t)\propto |P_s^{(3)}(t)|^2$.
By fulfilling a phase matching condition, we have that $|P^{(3)}(t)|=|P_s^{(3)}(t)|$.
Therefore, the signal is directly linked to an observable in the sense of quantum mechanics,
$
I_S(t)\propto |P^{(3)}(t)|^2
$  \cite{foot}.
Finally, we remark that the use of narrow pulses or the semi-impulsive limit approximation leads to
$
I_S(t)\propto |S(t,t_1,t_2)|^2
$. All these derivations are detailed in  \citep{Mukamel}.

\section{III. Generalized Quantum Invasiveness}

As previously discussed, it is always possible to perform measurements on classical systems with arbitrarily small disturbance on their evolutions. 
In turn, quantum mechanics allows disturbances on a physical system's evolution which cannot be minimized or gotten rid of classically.
Leggett and Garg acknowledged this sort of nonclassical disturbance with regards to the effect of measurements on the evolution of quantum systems \citep{LG}.
More generally, the effect of a general operation on the evolution of the system provide yet another route to witness nonclassicality as the quantum invasiveness of that operation \cite{Knee2,NewKnee, Moreira4}.
As we will see in detail, the quantum invasiveness of an operation can be unveiled by looking at the measurement statistics of an observable $Q$ subsequently measured and can be related to the presence of coherence in a certain basis.

\begin{figure}[h!]
\centering 
\includegraphics[width=7.8cm]{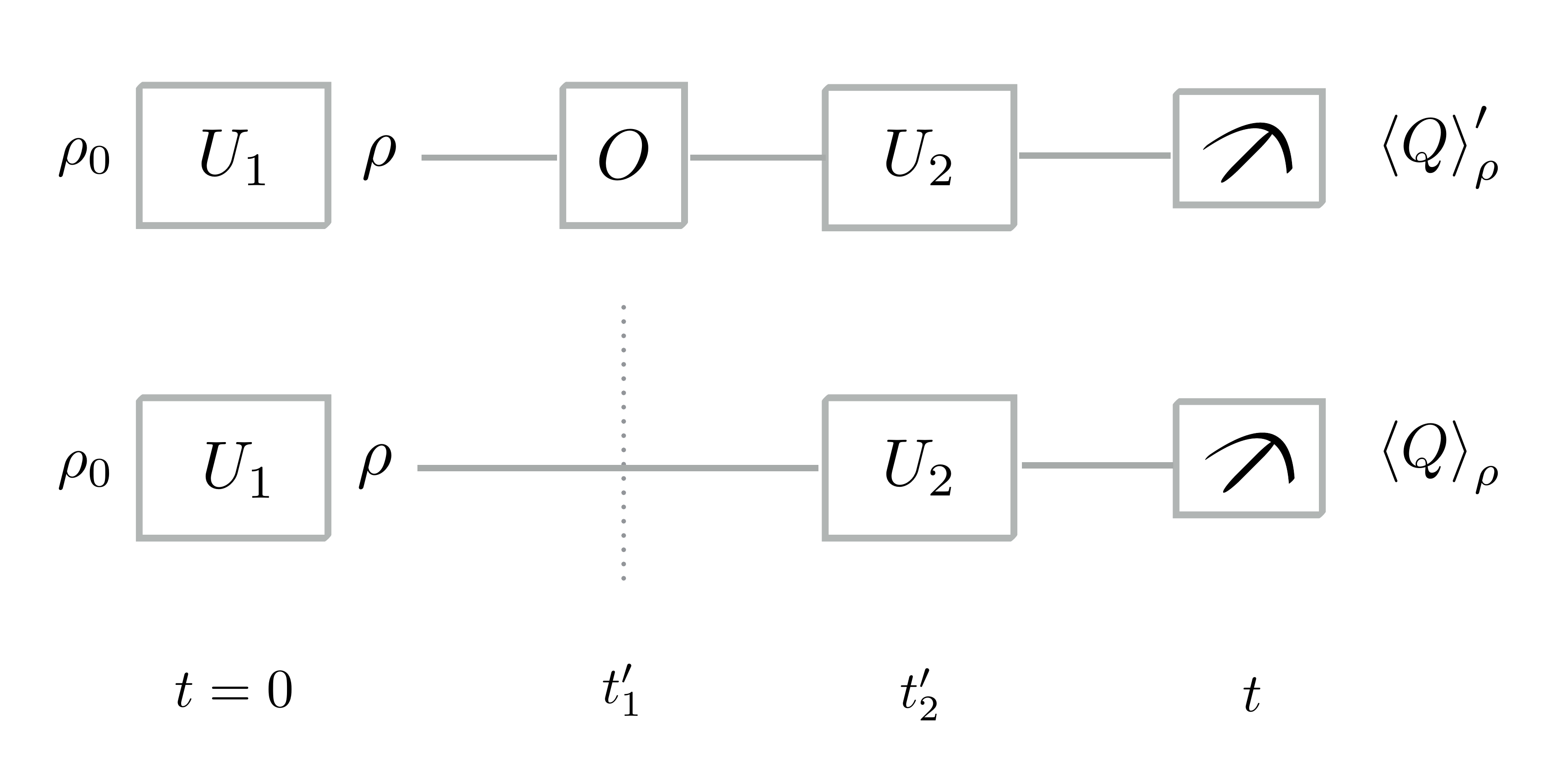}
\caption{ Scheme of the main experiment: in the upper part, a system initially in the state $\rho_0$ is subjected to the transformation $U_1$ at $t=0$ leading to state $\rho$. Then, an operation $O$ is applied at $t^\prime_1$. Finally, the measurement of the observable $Q$ is performed at $t^\prime_2$. The same realization is considered in the lower part of the figure, except for the fact that $O$ is not carried out at $t^\prime_1$.\label{Scenario1}}
\end{figure}

Let us consider the scheme proposed in Ref. \citep{Knee2} and sketched in Fig. \ref{Scenario1}.
This scheme will be presented for a two-level system but it can be easily extended for higher dimensions -- we will do this later in the context of the connection with spectroscopy. 
The kets $\{\ket{e},\ket{g}\}$ are the eigenstates of a chosen observable $\mathcal{O}$, not necessarily the measurement observable $Q$. 
They are referred to as \textit{classical states} in the sense that they are associated with definite values of that observable and therefore can be given a classical ontological interpretation.
In this sense, convex combinations of these eigenstates are also classical states.
In Fig. \ref{Scenario1}, we illustrate the preparation of the state $\rho$ as a result of an arbitrary state $\rho_0$ subject to a transformation represented by the box labeled as $U_1$.
For example, if $U_1$ is a unitary, then $\rho = U_1\rho_0 U^\dagger_1$.
This representation, which explicitly includes this first transformation, will be convenient for addressing the connection with spectroscopy experiments later.

\begin{figure}[h!]
\centering 
\includegraphics[width=7.8cm]{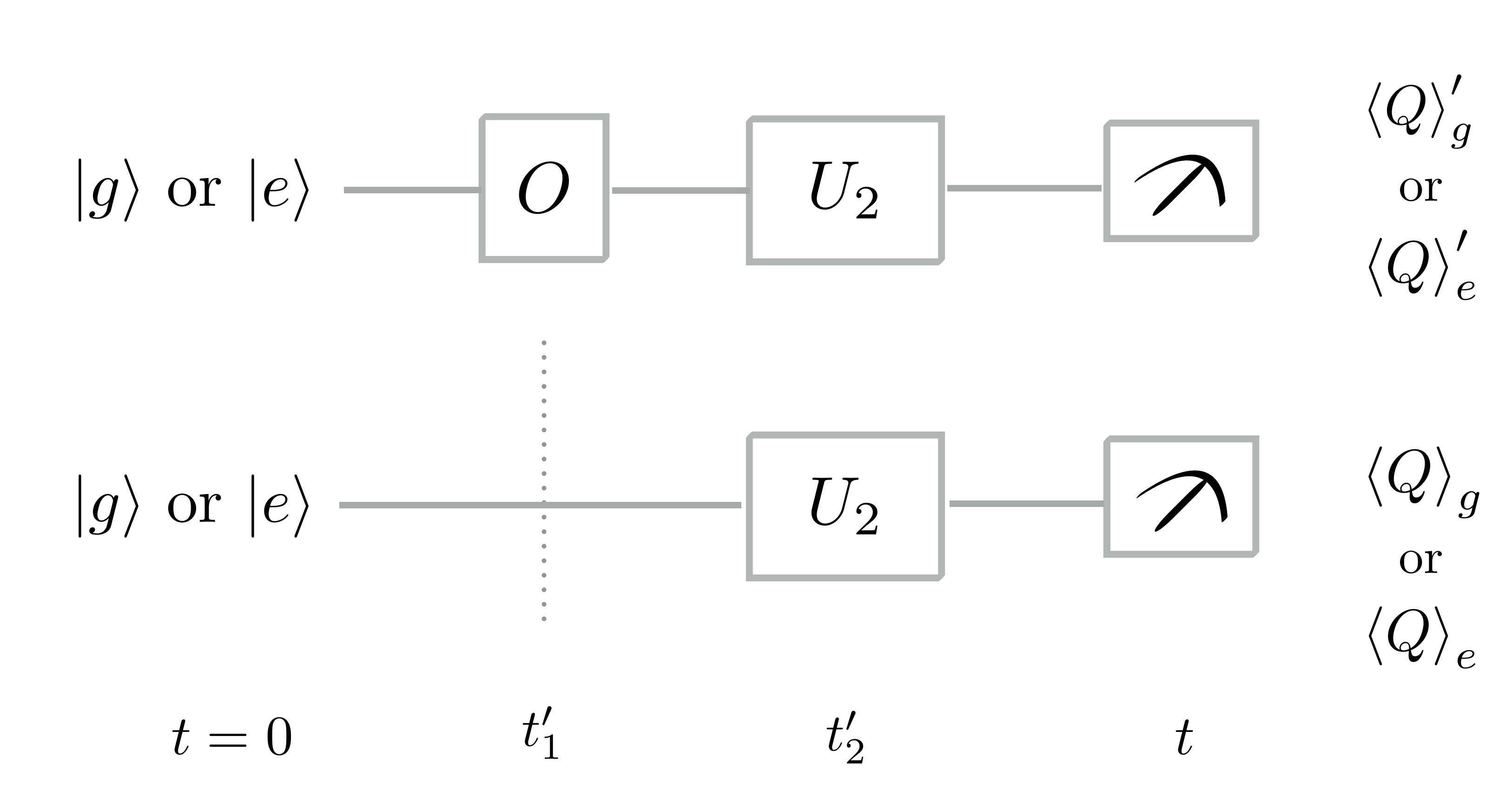}
\label{fig:fig14}
\caption{ Scheme of the control experiment: in the upper part, the classical states $\ket{g}$ and $\ket{e}$ prepared at $t=0$ are subjected to the operation $O$ at $t^\prime_1$. The measurement of the observable $Q$ is performed at $t^\prime_2$. The same realization is considered in the lower part of the figure, except for the fact that $O$ is not carried out at $t^\prime_1$\label{fig:control}}
\end{figure}

Now, the operation $O$ is performed at $t_1$, and the system undergoes a second transformation represented by the box labelled as $U_2$. 
Finally, the observable $Q$ is measured.
By repeating the experiment many times, one is able to obtain the expected value $\left< Q\right>^\prime_\rho$ (the $^\prime$ indicates the presence of the operation $O$). 
In the lower part of the figure, the experiment is now performed without the operation $O$. 
In this case, the expected value is denoted by $\left< Q\right>_\rho$. 
One then defines the following quantity
\begin{equation}\label{witness}
d_\rho\equiv \left<Q\right>^\prime_\rho-\left<Q\right>_\rho.
\end{equation}
The evaluation of this quantity will also be carried out in the so called control experiments \citep{Knee2} depicted in Fig.  \ref{fig:control} and defined as those which have as inputs  the classical states defined as the eigenstates of $\mathcal{O}$, i.e. either $\rho=\ket{g}\bra{g}$ or  $\rho=\ket{e}\bra{e}$.
In this way, one can define classical disturbances through  $d_g= \left<Q\right>^\prime_g-\left<Q\right>_g $ and 
$ d_e= \left<Q\right>^\prime_e-\left<Q\right>_e
$.
Nonclassical behaviour is then spotted if the following inequality is violated
\begin{equation}\label{inequality}
\min(d_g,d_e)\le d_\rho \le \max(d_g, d_e).
\end{equation}
Naturally, the violation of \eqref{inequality} can be related to the presence of coherence in the basis formed by the eigenstates of $\mathcal{O}$, as any convex combination of these eigenstates satisfy inequality \eqref{inequality} \citep{Knee2}.
These auxiliary experiments  serve for one fundamental purpose: it fixes the classical range for  $d_\rho$ in the main experiment.
This classical range can be enlarged due to the presence of classical disturbances which includes, for example, environmental noise and imprecisions in the application of the operation $O$.

It is important to note that nothing that has been said so far relies on the nature of operation $O$ at $t^\prime_1$.
Actually, $O$ can be a measurement operator, a unitary operation or even a general quantum channel \cite{NewKnee, Moreira4}.
To see this, consider $O$ as an arbitrary quantum operation.
One can therefore expect that, in general, \eqref{witness} will be trivially different from zero.
Nonetheless, the realization of the control experiments takes the eigenstates of $\mathcal{O}$ as inputs, as shown in Fig. \ref{fig:control}, which belongs to the set of classical states.
As we have seen above, the set of classical states includes the convex combinations of these eigenstates.
In turn, the control experiments will take into account the classical scenarios for $\ket{g}$ and $\ket{e}$ by determining the values  $d_g$ and $d_e$.
Then, by convexity, for any $\rho$ belonging to the set of convex combinations of these eigenstates, prepared as a result of the first operation $U_1$ in main the experiment (Fig. \ref{Scenario1}), inequality \eqref{inequality} will be satisfied.

Note that performing the control experiments with the classical states $\ket{g}$ and $\ket{e}$ as inputs is not the only way to determine  $d_g$ and $d_e$.
Alternatively, one may carry out this experiment with some classical state $\rho^C$ as input, an equilibrium Gibbs state, for instance, where
\begin{equation}\label{classicalState}
\rho^C=p_g\ket{g}\bra{g}+p_e\ket{e}\bra{e},
\end{equation}
and $p_g+p_e=1$. 
For this state as input, we have $d_{\rho^C}=p_g d_g + p_e d_e$.
Given that $p_g,p_e$ are known and by experimentally determining $d_{\rho_1^C}$ and $d_{\rho_2^C}$ with $\rho_1^C\neq\rho_2^C$, $d_g$ and $d_e$ can be evaluated.

\section{IV. Nonclassicality in spectroscopy experiments}

 We now establish a connection between disturbance and nonlinear electronic spectroscopy experiments.  
As discussed above, the interactions between the system and the laser pulses will be treated within the framework of the standard perturbative approach, which is the theoretical framework used to analyze these experiments \citep{Mukamel}. The connection will be established by identifying observables which are commonly measured in spectroscopy experiments and that can be considered as the measurement observable  $Q$ in the nonclassicality test explained in the previous section.
This represents a clear advantage from the point of view of the experimental viability.
Before going into the details, it is important to remark that there are alternative theoretical descriptions for electronic spectroscopy that could have been considered \citep{Cheng, Gelin}, and this can be investigated elsewhere.

We take $U_1$, $O$, and $U_2$ depicted in Fig. \ref{Scenario1} as the three laser pulses in the spectroscopy experiment depicted in Fig. \ref{2DES}. 
The first laser may drive the system out of equilibrium and produce the state $\rho$ depicted in  Fig. \ref{Scenario1}.
As the classical states, we will be considering the eigenstates of the multichromophoric system Hamiltonian, i.e.,  $\mathcal{O}=H_0$,  with $\ket{g}$ being the ground state and $\ket{i}$, with $i=1,2\ldots,n$, the excited states (exciton states). This choice will allow us to spot the presence of quantum coherence in the exciton basis, a topic of much interest to the quantum biology community \cite{Engel,Ishizaki,Panitchayangkoon}.

Consider first the top part of Fig.~\ref{Scenario1} of the main experiment. By starting with the state $\rho_0$, typically a Gibbs state, $\rho$ is prepared as the result of the first laser pulse $U_1$. 
The system then evolves freely under $H_0$ until the second pulse $O$ is applied. Once again,  free evolution takes place until the interaction with the last pulse $U_2$. 
The intensity of the emitted signal, $I^\prime_S(t)$, which is given in terms of the third-order polarization $P^{(3)}(t)$, is measured by a spectrometer at time $t$.
Consider now the lower part of Fig.~\ref{Scenario1}, where  the second laser pulse represented by $O$ is not applied. 
Only two interactions with the light pulses, $U_1$ and $U_2$, take place. As a result, the intensity  $I_S(t)$ is given in terms of the second-order polarization, $P^{(2)}(t)$. 
Hence,  Eq. \eqref{witness} can be expressed in terms of the induced polarizations as
\begin{equation}\label{PN}
d_\rho = I^\prime_S(t) - I_S(t) = |P^{(3)}(t)|^2 -  |P^{(2)}(t)|^2.
\end{equation}
If semi-impulsive limit is considered, one can instead define a related quantity
$
d^S_\rho = |S^{(3)}(t,t_1,t_2)|^2 -  |S^{(2)}(t,t_1,t_2)|^2.
$

We should also take into consideration the control experiments shown in Fig. \ref{fig:control}. 
A total of $n+1$ control experiments should be carried out, each of them corresponding to one of the eigenstates of $H_0$  as input, or different equilibrium states as explained before. As one can see, in the upper part of Fig. \ref{fig:control}, given one of the aforementioned eigenstates as input, only $O$ and $U_2$ are applied. In the lower part, in turn, only $U_2$ is carried out for the same eigenstate as input. 
As a result, we obtain $d_g= |P^{(2)}_g(t)|^2 - |P^{(1)}_g(t)|^2$ for the ground state and $d_i=|P^{(2)}_i(t)|^2 - |P^{(1)}_i(t)|^2$ for the excited states.
Consequently, the invasiveness witness of Eq. \eqref{inequality} now reads
\begin{equation}\label{Witness}
\min_i\{d_g,d_i\}\le d_\rho \le \max_i\{d_g,d_i\}.
\end{equation}
Therefore, we have found a sound connection between disturbance and electronic spectroscopy experiments, since  $d_\rho$ is given in terms of the polarization, an observable which is commonplace in the context of these experiments. 

As an illustrative example, let us consider an electronically coupled dimer with the Hamiltonian $(\hbar=1$)  \citep{Branczyk} 
\begin{equation}\label{Hamiltonian}
H_0=  \omega_A a_A^\dagger a_A + \omega_B a_B^\dagger a_B + J(a_A^\dagger a_B + a_B^\dagger a _A),
\end{equation}
where $ a_i^\dagger $ and $ a_i $ are, respectively, creation and annihilation operators for electronic excitations at the chromophore $ i \in  \{ A, B \} $, $ \omega_A $ and $ \omega_B $ are the first and second site energies, and $ J $ is the coupling between the chromophores. By defining the parameters $ \omega = (\omega_A + \omega_B)/2 $, $ \Delta = (\omega_A - \omega_B)/2$, and  $ \theta = \arctan(J/\Delta)/2 $, one can diagonalize the dimer Hamiltonian with the help of the transformed operators 
$
a_\alpha = \cos \theta a_A + \sin \theta a_B, 
$
and
$
a_\beta = -\sin \theta a_A + \cos \theta a_B
$
\citep{Cheng,Branczyk}. In this way, the Hamiltonian \eqref{Hamiltonian} becomes
\begin{equation}\label{Hamiltonian2}
H_0 = \hbar \omega_\alpha \ket{\alpha}\bra{\alpha} + \hbar \omega_\beta \ket{\beta}\bra{\beta} + \hbar \omega_f \ket{f}\bra{f},
\end{equation}
with the eigenvalues $\omega_\alpha = \omega + \Delta \sec 2 \theta, $ 
$\omega_\beta = \omega - \Delta \sec 2\theta, $ and
$\omega_f = \omega_a + \omega_b.$
Without any loss of generality, the ground state $ \ket{g} $ is associated with the eigenvalue $\omega_g=0$. In \eqref{Hamiltonian2}, we also have $ \ket{\alpha} = a_\alpha^\dagger \ket{g} $ and $ \ket{\beta} = a_{\beta}^\dagger \ket{g} $  as single exciton states and $ \ket{f} = a_\alpha^\dagger a_\beta^\dagger \ket{g} $ the two-exciton state.
The energy eigenstates of the system, $ \{ \ket{g}, \ket{\alpha}, \ket{\beta},  \ket{f} \} $, constitute the exciton basis and are the classical states in our protocol to witness nonclassicality. 

By assuming inversion symmetry, the even-order nonlinear polarizations must vanish \citep{Boyd}.
Hence, $d_\rho = |P^{(3)}(t)|^2$ and, for the control experiments, $d_j = -|P^{(1)}_j(t)|^2$ with $j = g, \alpha, \beta,f$.
The maximum value $d_j$ can assume is, therefore, $d_j =0$.  
In this case, according to inequality \eqref{Witness}, quantum invasiveness would in principle be witnessed whenever
\begin{equation}\label{FinalWitness}
|P^{(3)}(t)| \neq 0.
\end{equation}
Therefore, for the particular case where $P^{(2)}(t)=0$, $|P^{(3)}(t)|\neq 0$ implies nonclassicality.
Moreover, if the semi-impulsive limit applies, \eqref{FinalWitness} can be written as $S^{(3)}(t,t_1,t_2)\neq 0$. In this example, one can see that the \textit{whole} protocol, including both the main and control experiments, were reduced to one single condition: $|P^{(3)}(t)|\neq 0$. 
Note, however, that this condition assumes the absence of underlying experimental errors associated with the particular experimental arrangement.
Nonetheless, the control experiments have the purpose to address the clumsiness loophole \citep{Wilde} as well.
In our context, the clumsiness loophole means the possibility of explaining the violation of inequality \eqref{Witness} classically.
 It can be viewed, therefore, as a result of errors and classical disturbance.
As mentioned above, it can be addressed by considering classical scenarios, which are a subclass of the more general class of scenarios represented in Fig. \ref{Scenario1} \citep{Knee2, NewKnee, Moreira4}.
Classical scenarios correspond to the experiments which have a classical state as input, the classical states being defined as the eigenstates of the observable $\mathcal{O}$, as well as convex mixtures of these eigenstates -- examples of such classical scenarios are the control experiments shown in Fig. \ref{fig:control}.
However, as discussed before, these control experiments can be carried out by using, for example, equilibrium Gibbs states as inputs to determine $d_g$ and the $d_i$'s.
Through many realizations of the control experiments, one is able to statistically detect errors and possible classical disturbances associated with the particular experimental arrangements by evaluating $d_g$ and the $d_i$'s.
Then, we see that our protocol provides a way to block the clumsiness loophole since it allows one to statistically detect classical disturbances, reflecting on the bounds of inequality \eqref{Witness}. We stress that such errors, including not exactly vanishing two-order nonlinear polarizations and different noise mechanisms associated with the particular experimental setup, can only be detected when the experiment is ultimately performed.
Inequality \eqref{Witness} will be used to take them into account even when $d_g=d_i=0$ is obtained theoretically, as it is the case for the result expressed in \eqref{FinalWitness}.
As a result, by minimizing experimental errors and noise, one increases the possibility of witnessing nonclassicality.

For completeness, let us explicitly include some noise to the electronic dimer, what will undoubtedly make the example a bit more realistic. However, we will still keep only electronic states in the system description as vibrational or vibronic effects would render this first discussion unnecessary more involved. This does not mean that vibrational degrees of freedom are not present. They are exactly the source of the environmental noise we are going to discuss below. What we will not do here is to consider dimers or other molecular structures where the coupling to vibration is so intense that this degree of freedom must be included in the system Hamiltonian. We are considering them as part of the bath or environment for the electronic part. An import source of noise is dephasing. This causes the coherences to progressively vanish while keeping the populations unchanged. Physically, the vibration and rotation of the molecule will cause local changes in the charge distribution over the molecule. As a result, the local frequencies $\omega_A$ and $\omega_B$ of the dimer will be stochastically changed by the incoherent thermal motion of external degrees of freedom. Typically, this is described by augmenting  Hamiltonian in Eq. \eqref{Hamiltonian2} with terms concerning a bosonic bath and its coupling to the dimer \citep{Li}.
By assuming a weak coupling to the bath, what is consistent with the fact that the vibrational degrees of freedom are not explicitly considered as part of the dimer, one can deduce a master equation for the system density matrix $\sigma(t)$ in between the pulses \citep{Mukamel,Li}
\begin{equation}\label{Noise}
\dot{\sigma}_{\nu\nu^\prime} = -\frac{i}{\hbar}(\epsilon_\nu-\epsilon_{\nu^\prime})\sigma_{\nu\nu^\prime} - \Gamma_{\nu\nu^\prime}\sigma_{\nu\nu^\prime},
\end{equation}
with  $\nu,\nu^\prime=g,\alpha,\beta,f$ and $\epsilon_\nu,\epsilon_{\nu^\prime}$ the eigenvalues associated with the eigenstates $\ket{\nu}$ in the exciton representation. Also, $\Gamma_{\nu\nu^\prime}$ contains the relaxation rates.
For simplicity, we take all the relaxation rates to be the same, $\Gamma_{\nu\nu^\prime}=\Gamma$ and assume the semi-impulsive limit.



Let us now consider how dephasing affects the result presented before. Clearly, in between the application of the pulses, the diagonal elements of the system density matrix die out according to \eqref{Noise}. This affects the measured signal through $S^{(3)}(t,t_1,t_2)\propto e^{-\Gamma t}$  \citep{Mukamel, Hamm}.
The practical consequence for the protocol presented here is precisely what has been discussed before as the role of generic errors. As the third-order response function will decay exponentially and tend to zero, this implies that \eqref{FinalWitness} becomes less likely to be verified as time increases.
Therefore, we see that the presence of noise in the Lindblad form decreases the value of $d_\rho$.
As a conclusion, it affects even more the possibility of witnessing nonclassicality in the presence of experimental errors or imprecisions, i.e. when condition \eqref{FinalWitness} will not be valid anymore. 
Instead, inequality \eqref{Witness} which eventually takes classical disturbances and environmental noise into account in its bounds, which should be ultimatelly determined through the control experiments, must be violated.


\section{V. Final considerations}

Violations of the LGI for the multichromophoric systems were found theoretically in Ref. \citep{LGWilde}, where projective measurements in the site-basis were employed.
However, as pointed out by the authors, such measurements are not realistic when it comes to experimental implementations. Our proposal aims to bridge this gap by linking nonclassicality tests to experimental techniques in a direct and fundamental way. By nonclassicality, in the scope of our work, we mean the nonclassical effect of an operation on the statistics of an observable subsequently measured. This can be associated both with coherence in the basis of an observable and the lack of an ontological interpretation of the measurement results given the deviation from convex sums of classical results, i.e., state superposition.
Additionally, given that our work investigate quantum invasiveness in driven systems \citep{Friedenberger}, 
it deals with setups which are not conventionally considered in Leggett-Garg scenarios.
At the same time, by proposing sound protocols which employ well-established experimental techniques such as electronic spectroscopy, we expect that the present work will pave the way for new and exciting experimental investigations in quantum biology and quantum technologies that rely on time-dependent Hamiltonians.

\textit{Acknowledgments} - We thank Mauro Paternostro for valuable discussions. S. V. M. acknowledges financial support from the Brazilian agency CAPES and F. L. S.
acknowledges partial support from of the Brazilian National
Institute of Science and Technology of Quantum Information
(INCT-IQ) and CNPq (Grant No. 302900/2017-9).


\begin{thebibliography}{99}

\bibitem{Mukamel} S. Mukamel, \textit{Principles of Nonlinear Optical Spectroscopy} (Oxford University
Press, 1995).
\bibitem{pathways} Y. -C. Cheng, G. S. Engel, G. R. Fleming, Chem. Phys. \textbf{341}, 285 (2007).
\bibitem{Lambert} N. Lambert, Y.-N. Chen, Y.-C. Cheng, C.-M. Li, G.-Y. Chen and F. Nori, Nat. Phys. {\bf 9}, 10 (2013).
\bibitem{Engel} G. S. Engel, T. R. Calhoun, E. L. Read T.-K. Ahn, T. Mancal and Y.-C. Yuan, R. E. Blankenship and G. R. Fleming, Nat. {\bf 446}, 782 (2007).
\bibitem{Ishizaki} A. Ishizaki and G. R. Fleming, Proc. Nat. Acad. Sc. {\bf 106}, 41 (2009).
\bibitem{Panitchayangkoon} G. Panitchayangkoon, D. Hayes, K. A. Fransted, J, R. Caram, E. Harel, J. Wen, R. E. Blankenship and G. Engel, Proc. Nat. Acad. Sc. {\bf 107}, 12766 (2010).
\bibitem{Collini} E. Collini, C. Y. Cathy, K. E. Wilk, P. M. Curmi, P. Brumer and G. D. Scholes, Nat. {\bf 463}, 644 (2010).
\bibitem{LG} A. J. Leggett  and A. Garg, Phys. Rev. Lett. {\bf 54}, 857 (1985).
\bibitem{Emary} C. Emary, N. Lambert, F. Nori, Rep. Prog. Phys. {\bf 77}, 016001 (2014).
\bibitem{Kofler2} J. Kofler and C. Brukner, Phys. Rev. A {\bf 87}, 052115 (2013).
\bibitem{Moreira} S. V. Moreira, A. Keller, T. Coudreau and P. Milman, Phys. Rev. A {\bf 92}, 062132 (2015).
\bibitem{Clemente} L. Clemente and J. Kofler, Phys. Rev. A {\bf 91}, 062103 (2015).
\bibitem{Maroney} O. J. E. Maroney and C. G. Timpson, arXiv: 1412.6139v1 (2014).
\bibitem{Kumari} S Kumari and A. K. Pan, EPL {\bf 118}, 50002 (2017).
\bibitem{Clemente2} L. Clemente and J. Kofler, Phys. Rev. Lett. {\bf 116}, 150401 (2016).
\bibitem{Martin} A. Smirne, D. Egloff, M. G. D\'iaz, M. B. Plenio, and S. F. Huelga,  Quantum Sci. Technol. {\bf 4}, 01LT01 (2019).
\bibitem{Animesh} G. C. Knee, M. Marcus, L. D. Smith, and A. Datta, Phys. Rev. A {\bf 98}, 052328 (2018).
\bibitem{Knee2} G. C. Knee, K. Kakuyanagi, M-.C. Yeh, Y. Matsuzaki, H. Toida, H. Yamaguchi, A. J. Leggett and W. J. Munro, Nature Comm. {\bf 7}, 13253 (2016).
\bibitem{Schild} G. Schild and C. Emary, Phys. Rev. A {\bf 92}, 032101 (2015).
\bibitem{Lambert2} C.-M. Li, N. Lambert, Y.-N. Chen, G.-Y. Chen and F. Nori, Scientific Reports {\bf 2}, 885 (2015).
\bibitem{NewKnee} K. Wang, G. C. Knee, X. Zhan, Z. Bhian, J. Li and P. Xue, Phys. Rev. A {\bf 95}, 032122 (2017).
\bibitem{Moreira4} S. V. Moreira and M. Terra Cunha, Phys. Rev. A {\bf 99}, 022124 (2019).
\bibitem{foot} The macroscopic polarization $P$ corresponds to an ensemble average of the quantum mechanical electric dipole operator  \citep{Mukamel}.
\bibitem{Gelin} M.F. Gelin, D. Egorova, W. Domcke, J. Chem. Phys. {\bf 123}, 164112 (2005).
\bibitem{Cheng} Y.-C. Cheng, G. S. Engel and G. R. Fleming, Chem. Phys. {\bf 341}, 285 (2007).
\bibitem{Branczyk} A. M. Branczyk, D. B. Turner and G. D. Scholes,  Ann. Phys. {\bf 526}, 31 (2014).
\bibitem{Boyd} R. W. Boyd, \textit{Nonlinear optics} (Academic Press, 2008).
\bibitem{Wilde} M. M. Wilde and A. Mizel, Found. Phys. {\bf 42}, 256 (2012).
\bibitem{Hamm} P. Hamm, Principles of Nonlinear Optical Spectroscopy: A Practical Approach (2005).
\bibitem{LGWilde} M. M. Wilde, J. M. Mccraken and A. Mizel, Proc. R. Soc. A {\bf 466}, 1347 (2010).
\bibitem{Li} H.-r. Li, P. Zhang, Y. Liu, F.-l. Li and S.-y. Zhu, Phy. Rev. A {\bf 87}, 053831 (2013).
\bibitem{Friedenberger} A. Friedenberger and E. Lutz, arXiv: 1805.11882 (2018).





\end{thebibliography}
\end{document}